\documentclass[useAMS,usenatbib]{mn2e}
\usepackage{epsfig,graphicx,lscape}
\title[IVCs and HVCs towards POP early-type stars]{Observations towards 
early-type stars in the ESO-POP survey: II -- 
searches for intermediate and high velocity clouds}

\author[J. Smoker et al.]
         {J.~V. Smoker$^{1}$\thanks{email: j.smoker@qub.ac.uk. \hspace*{0.3cm} 
          \newline Based on 
          observations taken at UT2, Kueyen, Cerro Paranal, Chile, 
           ESO DDT programme 266.D-5655(A), UVES Paranal Observatory Project, with additional 
	   observations from 071.B-0529(A), 072.B-0585(A), 073.B-0607(A), 074.B-0639(A), 076.D-0018(A) and 077.D-0025(A).},
          I. Hunter$^{1}$, P.~M.~W Kalberla$^{2}$, F.~P. Keenan$^{1}$, R. Morras$^{3}$, \newauthor
	  R. Hanuschik$^{4}$, H.~M.~A. Thompson$^{1}$, 
	  D. Silva$^{5}$, E. Bajaja$^{3}$, W.~G.~L Poppel$^{3}$, \newauthor
	  M. Arnal$^{3}$
          \\
          \\
       $^{1}$Astrophysics Research Centre,
            Department of Physics and Astronomy,
            Queen's University Belfast, \\
            Belfast, BT7 1NN,
            U.K.                         \\
       $^{2}$Argelander-Institut f\"ur Astronomie, Universit\"at
             \thanks{Founded by merging of the
             Sternwarte, Radioastronomisches Institut, and Institut f\"ur Astrophysik
             und Extraterrestrische Forschung der Universit\"at Bonn.},
             Auf dem H\"ugel 71, 53121 Bonn, Germany \\
       $^{3}$Instituto Argentino de Radioastronomia,
            Casilla de correo 5,
            Villa Elisa,
            Argentina.                   \\	
       $^{4}$European Southern Observatory,
             Karl-Schwarzschild-Str. 2, 
	     D-85748 Garching bei München,
	     Germany                     \\
       $^{5}$TMT Observatory Scientist,
             AURA/Thirty Meter Telescope,
	     2636 East Washington Blvd.,
	     Pasadena, CA 91107,
             U.S.A.   \\ 	 
}

\date{Accepted
      Received
      in original form }

\def\LaTeX{L\kern-.36em\raise.3ex\hbox{a}\kern-.15em
    T\kern-.1667em\lower.7ex\hbox{E}\kern-.125emX}

\pubyear{}

\begin{document}
\label{firstpage}
\maketitle
\begin{abstract}

We present Ca\,{\sc ii} K and Ti\,{\sc ii} optical spectra of early-type 
stars taken mainly from the UVES Paranal Observatory Project, plus
H\,{\sc i} 21-cm spectra from the Vila-Elisa and Leiden-Dwingeloo surveys, 
which are employed to obtain distances to intermediate and high velocity clouds (IHVCs). 
H\,{\sc i} emission at a velocity of --117 km\,s$^{-1}$ towards 
the sightline HD\,30677 ($l,b$=190.2$^{\circ}$,--22.2$^{\circ}$) with column 
density $\sim$1.7$\times$10$^{19}$ cm$^{-2}$ has no corresponding Ca\,{\sc ii} K 
absorption in the UVES spectrum, which has a signal-to-noise (S/N) ratio of 
610 per resolution element. The star has a spectroscopically determined distance of 2.7-kpc, 
and hence sets this as a firm lower distance limit towards Anti-Centre cloud ACII. Towards another sightline 
(HD\,46185 with $l,b$=222.0$^{\circ}$, --10.1$^{\circ}$), H\,{\sc i} at a velocity of +122 km\,s$^{-1}$ and column density of 
1.2$\times$10$^{19}$ cm$^{-2}$ is seen. The corresponding Ca\,{\sc ii} K spectrum has a S/N = 780,
although no absorption is observed at the cloud velocity. This similarly places a firm lower distance limit of 
2.9-kpc towards this parcel of gas that may be an intermediate velocity cloud. The lack of intermediate velocity (IV) 
Ca\,{\sc ii} absorption towards HD\,196426 ($l,b$=45.8$^{\circ}$,--23.3$^{\circ}$) at a S/N = 500 
reinforces a lower distance limit of $\sim$700-pc towards this part of Complex gp, where the H\,{\sc i} 
column density is 1.1$\times$10$^{19}$ cm$^{-2}$ and velocity is +78 km\,s$^{-1}$. Additionally, no IV Ca\,{\sc ii} is 
seen in absorption in the spectrum of HD\,19445, which is strong in H\,{\sc i} with a column density of 8$\times$10$^{19}$ cm$^{-2}$ at a 
velocity of $\sim$--42 km\,s$^{-1}$, placing a firm although uninteresting lower distance limit of 39-pc to this part of IV South. 
Finally, no HV Ca\,{\sc ii} K absorption is seen towards HD\,115363 ($l,b$=306.0$^{\circ}$,--1.0$^{\circ}$) at a S/N = 410, 
placing a lower distance of $\sim$3.2-kpc towards the HVC gas at velocity of $\sim$+224 km\,s$^{-1}$ and H\,{\sc i} column 
density of 5.2$\times$10$^{19}$ cm$^{-2}$. This gas is in the same region of the sky as complex WE (Wakker 2001), but 
at higher velocities. The non-detection of Ca\,{\sc ii} K absorption sets a lower distance of $\sim$3.2-kpc towards the 
HVC, which is unsurprising if this feature is indeed related to the Magellanic System. 

\end{abstract}


\section{Introduction}                                           \label{s_intro}

The Paranal Observatory Project (POP; Bagnulo et al. 2003)\footnote{See also http://sc.eso.org/santiago/uvespop/} provides a wealth of
high-resolution ($R\sim$80,000) optical spectra towards stars mainly in the Galactic disc 
that can be used to study subjects such as stellar properties, kinematics and the interstellar
medium. In a previous paper (Hunter et al. 2006, hereafter Paper {\sc i}) we used a 
sample of early-type stars in the POP survey in order to investigate the 
interstellar medium in the Na\,{\sc i} UV, Ti\,{\sc ii} and Ca\,{\sc ii} K lines,
using the stars as light sources to probe the material between the star and the 
Earth. Because these O- and B-type stars are often fast rotators with weak 
metal lines, they are ideal for probing narrow interstellar features. 

In the current paper, we use mainly Ca\,{\sc ii} and Ti\,{\sc ii} spectra 
in order to search for Intermediate and High Velocity Clouds (IHVCs) towards 
the sightlines investigated in Paper {\sc i}, plus some additional sightlines for which 
high S/N data have now become available. Our aim is to improve the distances to 
these still enigmatic objects by searching for IHVCs in the Villa-Elisa
Southern Sky 21-cm H\,{\sc i} Survey (Bajaja et al. 2005) or Northern 
Hemisphere counterpart, the Leiden-Dwingeloo Survey (Hartmann \& Burton 1997), and
subsequently searching for corresponding absorption in the Ca\,{\sc ii} or Ti\,{\sc ii} optical 
spectra. 
Although the distance to many Intermediate Velocity Clouds  (IVCs) is known 
(e.g. Kuntz \& Danly 1996 and references therein), to date there are very few uncontroversial 
upper distance limits towards High Velocity Clouds (HVCs). Indeed towards the main complexes there
are currently only three uncontroversial detections; towards Complex A (van Woerden et al. 1999), 
Complex M (Danly, Albert \& Kuntz 1993) and Complex WB (Thom et al. 2006).  Hence it is still unclear whether
many of these objects are associated with the Galaxy, for example linked 
with a Galactic fountain with distances of $\approx$ 10 kpc (e.g. Quilis \& Moore 2001), or 
are failed dwarf galaxies with distances of several hundred kpc (e.g. Braun \& Burton 1999).
Clearly, as the sample stars in the POP survey were not chosen $a-priori$ to intersect with 
known IHVC complexes, the majority of the sightlines do not cross known IHVCs. However, 
serendipitously a few of the sightlines intersect complexes and are 
studied in the current paper. In addition to the POP data, we include spectra 
from two recent high spectral resolution observing runs taken using the \'{e}chelle 
spectrometer FEROS, plus further UVES observations whose primary aim was to obtain spectra 
for a stellar library but that are also at high S/N and cover the Ca\,{\sc ii} K line (Silva et al. 2007).  

The current work complements our previous studies in which we obtained improved distance 
limits towards IVC complexes gp and K and HVC complexes C, WA-WB, WE, and H (Smoker et al. 2004, 2006), 
by searching for absorption in high-resolution spectra of mainly B-type stars taken from the 
Edinburgh-Cape (Stobie et al. 1997) and Palomar-Green Surveys (Green, Schmidt \& Liebert 1986). 
In particular the current sightlines intersect IVC Complex K with previous distance limit of 
0.7--6.8-kpc (de Boer \& Savage 1983, Smoker 2006), the Anti-Centre clouds with previous distance 
limit of $>$0.4-kpc (Tamanaha 1996) and Complex WE with distance limit $<$12.8-kpc 
(Sembach et al. 1991). Finally, one of our current sightlines lies towards the M\,15 intermediate 
velocity cloud lying in the IVC complex gp. This cloud has been studied extensively, in the optical to determine 
variations in velocity and equivalent width variations (Lehner et al. 1999, Meyer \& Lauroesch 1999, 
Smoker et al. 2002), plus in the H\,{\sc i}, infrared and H$\alpha$ (Kennedy et al. 1998, Smoker et al. 2002). 
An improvement in the current distance limit of 0.8--4.3 kpc (Wakker 2001 and references therein) would 
be very useful to more accurately define the cloud parameters such as cloudlet sizes and densities and 
to provide clues to the high metalicity of this IVC (Little et al. 1994).

Sect. \ref{sample} describes the sample, provides a table noting the 
cases where the current sightlines cross known IHVC complexes plus new observations 
not previously described in Paper I, and shows the 
optical and H\,{\sc i} spectra. Sect. \ref{results} gives the main results, 
including the cases where the current optical sightlines intersect 
IHVCs and an attempt to obtain improved distance limits 
towards these clouds. Sect. \ref{discussion} discusses the most interesting lower limits to IHVCS and 
finally Sect. \ref{summary} gives a  
summary of the main findings.

\section{The sample, observations and data reduction}          
\label{sample}

The list of sample stars is shown in Table \ref{t_stars}. The table includes all stars 
for which new observations were taken, plus sightlines that lie towards IHVC complexes 
that are discussed in Sect. \ref{distlim}, but does not include the POP paper {\sc i} 
objects that have no IHVC detection. Further information
concerning the POP objects is given in Paper {\sc i}. They are all O- and B-type stars 
with 2.3$<$ $m_{v}$ $<$7.9 mag. For these POP  
optical spectroscopic data, we used the on-line versions of reduced data from 
the Paranal Observatory Project (Bagnulo et al. 2003). These are spectra 
taken with the UVES \'{e}chelle spectrometer mounted on the 8.2-m Kueyen telescope
at the Very Large Telescope at a spectral resolution of 80,000 or 3.75 km\,s$^{-1}$ 
and S/N pixel$^{-1}$ ranging from 190--770. In this 
paper we concern ourselves with the Ca\,{\sc ii} K ($\lambda_{\rm air}$=3933.66\AA)
and Ti\,{\sc ii} ($\lambda_{\rm air}$=3383.76\AA) species only. A further 9 stars were 
observed with FEROS on the ESO 2.2-m on La Silla during observing sessions in Oct. 
2005 (FER1 in Table \ref{t_stars}) and May 2006 (FER2 in Table \ref{t_stars}). These 
stars are all B-type post-AGB stars or Planetary Nebulae and have fainter magnitudes than 
the POP stars, with 9.4$<$ $m_{v}$ $<$ 13.3 mag. The S/N ratios pixel$^{-1}$ at Ca\,{\sc ii} K range from 
$\sim$40--120 and the resolution is 
$R$=48,000. The spectra shown in this paper are the 
quick-look pipeline products. As a check of their reliability, during each of the FEROS 
runs a bright B-type star from the POP survey was observed and the velocities and equivalent 
widths of some of the absorption lines were compared between the two datasets. Agreement was 
found to be excellent. Finally, 12 stars were taken from the dataset of Silva et al. (2007; S07 in Table \ref{t_stars}). 
These are UVES spectra of early-type stars with 5.9 $<$ $m_{V}$ $<$ 11.3 mag., observed at a 
spectral resolution of $\sim$40,000 with S/N = 100--620 pixel$^{-1}$, and 
were 
reduced using the ESO pipeline (MIDAS context) with calibrations taken the morning after the observations. 
For the H\,{\sc i} 21--cm spectra, we used either the Southern Villa-Elisa H\,{\sc i} survey data (Bajaja et al. 2005), 
corrected for the effects of stray radiation or the Leiden-Dwingeloo survey for sightlines with Dec.$>$--20$^{\circ}$ 
(Hartman \& Burton 1997). Both surveys have been merged to form the
Leiden/Argentine/Bonn (LAB) H\,{\sc i} line survey (Kalberla et al. 2005) which has a velocity resolution of 1 km\,s$^{-1}$ and 
brightness temperature sensitivity of 0.07 K. 

In Table \ref{t_stars} the columns are as follows. 
Columns 1--5 give the star HD name, alternative name, Galactic coordinates and $V$-band magnitude
taken from {\sc simbad}. Columns 6--7 give the estimated stellar distance 
and $z$-height above or below the Galactic plane. These distances were primarily 
estimated using the method of spectroscopic parallax from the spectral type,
apparent magnitude and reddening towards each star, estimated from the observed 
$(B-V)$ colour. Absolute magnitudes as a function of spectral type were taken from Schmidt-Kaler (1982) with 
colours from Wegner (1994). 
Details are given in Paper {\sc i}. Excluding perhaps large systematic
errors caused by the uncertainty in the absolute magnitude calibration of our sample, the
distances have an uncertainty of $\sim$30 per cent. For a number of 
objects (in particular the Wolf-Rayet stars, peculiar objects and Post-AGB stars), distances were taken from 
the reference given at the foot of the table. For example for HD\,179407 the distance is given as 
7600$^{1}$ where the suffix refers to reference number 1 where the distance of 7600-pc was given. 
Column 8 gives the signal-to-noise (S/N) ratio pixel$^{-1}$ in the Ca\,{\sc ii} spectrum; to obtain 
the S/N per resolution element this needs to be multiplied by $\sqrt{2}$. 

If the coordinate of the 
sightline lies within any of the figures of Wakker (2001) which display 
H\,{\sc i} column densities towards IHVCs, this name is given in Column 9.
We must stress that although more than 35 of our stars lie within the 
boundaries of these figures, often they are in regions where no IHVC is 
observed in H\,{\sc i}, for example because the stars lie in holes in the H\,{\sc i} 
distribution. Columns 10 and 11 give the minimum and 
maximum expected LSR velocity for gas orbiting the Galactic Centre, 
based on the direction of the sightline
and the distance to the stellar target. To calculate the velocity range for "normal" gas, we use
the methodology of Wakker (1991), in that we assume a flat rotation curve
with $v$$_{\rm rot}$~=~220~km~s$^{-1}$ at $r>$0.5 kpc, decreasing linearly
towards the Galactic Centre, together with equations from Mihalas \&
Binney \cite{mih81}.  A deviation velocity for interstellar cloud
components which lie outside the expected velocity range is calculated,
where the deviation velocity is defined as the difference between the velocity 
of the component and the nearest limit of the expected velocity range 
(Wakker 1991). We classify low velocity clouds (LVCs) as having absolute values of their deviation 
velocities below 30~km~s$^{-1}$, IVCs between 30~km~s$^{-1}$ and 90~km~s$^{-1}$, and HVCs
greater than 90~km~s$^{-1}$. Finally, column 12 gives the source for the optical 
spectra (POP for stars from Paper {\sc i}; FER1/FER2 for FEROS observations; S07 for stars from 
Silva et al. 2007), and H\,{\sc i} data 
(LD for Leiden-Dwingeloo; VE for Villa-Elisa Survey sightlines).

\begin{table*}
\begin{center}
\small
\caption[]{The stellar subsample for new observations plus all sightlines which lie in the vicinity of IHVCs. The S/N ratios per pixel 
are for Ca\,{\sc ii} K (3933\AA). See text for details.}
\label{t_stars}
\begin{tabular}{lrrrrrrrrrrrr}
\hline
    Star        &    Alt.      &     $l$~~&    $b$~~ & $m_{v}$ &      $d$      &  $z$        &S/N            &  IHVC  & v$_{\rm dev}^{\rm min}$ & v$_{\rm dev}^{\rm max}$ & Source	     \\
                &   Name       &   (deg.) &   (deg.) &  (mag.) &     (pc)      &  (pc)       &pixel$^{-1}$   &        &                       &                	  & Opt./H\,{\sc i}  \\
\hline

HD\,171432       &  BD-18 5008 &   14.62  &   -4.98  &   7.11  &	4014	    &  -348  & 590 & --      &  0.0  & 43.9 & POP/VE   \\
EC\,20485-2420   &             &   21.76  &  -36.36  &  11.77  &      3600$^{5}$    & -1200  &  40 & gp      &  0.0  & 28.9 & FER1/VE  \\
HD\,179407       &  BD-12 5308 &   24.02  &  -10.40  &   9.44  &      7600$^{1}$    & -1400  & 120 & gp      &  0.0  &128.3 & FER1/VE  \\
HD\,188294       &  57 Aql B   &   32.65  &  -17.77  &   6.44  &	 212	    &	-64  & 420 & gp      &  0.0  &  2.3 & POP/VE   \\
G169--28         &  HIP 82398  &   41.83  &  +36.06  &  11.26  &        117$^{2}$   &    69  & 100 & K       &  0.0  &  1.0 & S07/LD   \\
HD\,196426       &  HR 7878    &   45.81  &  -23.32  &   6.21  &        700$^{3}$   &  -280  & 360 & gp      &  0.0  &  3.3 & S07/LD   \\
HD\,344365       &             &   58.63  &   +3.41  &  10.8   &       1032$^{13}$  &    61  & 210 & --      &  0.0  & 11.4 & S07/LD   \\
HD\,2857         &             &  110.05  &  -67.64  &   9.95  &        717$^{7}$   &  -663  & 270 & IVS     & -0.9  &  0.0 & S07/LD   \\
HD\,19445        &             &  157.48  &  -27.20  &   8.05  &         39$^{12}$  &   -18  & 200 & IVS, ACC& -0.3  &  0.0 & S07/LD   \\
HD\,30677        &  BD+08 775  &   190.18 &  -22.22  &   6.84  &	2707	    & -1023  & 430 & ACII      &0.0  &  8.1 & POP/VE   \\
HD\,46185        &  BD-12 1520 &  221.97  &  -10.08  &   6.79  &	2937	    &  -514  & 550 & --      &  0.0  & 31.1 & POP/VE   \\
BD-12\,2669      &             &  239.12  &  +18.17  &  10.22  &        158$^{8}$   &    49  & 250 & IV Spur &  0.0  &  1.6 & S07/LD   \\
HD\,72067        &  HR 3356    &  262.08  &   -3.08  &   5.83  &	 488	    &	-26  & 450 & --      &  0.0  &  2.0 & POP/VE   \\
EC\,05229-6058   &             &  269.97  &  -34.08  &  11.4   &       2200$^{5}$   & -2100  & 150 & --      &  0.0  &  4.1 & FER1/VE  \\
HD\,94910        &  HIP 53461  &   289.18 &   -0.69  &   7.09  &	6000$^{4}$  &	-72  & 430 & --      & -12.2 &  3.4 & POP/VE   \\
EC\,01483-6806   &             &  294.73  &  -48.36  &  11.1   &       2600$^{5}$   & -2000  & 130 & --      & -9.5  &  0.0 & FER1/VE  \\   
LB\,3193         &             &  297.32  &  -54.90  &  12.70  &       8000$^{6}$   &  1800  & 100 & --      & -14.4 &  0.0 & FER1/VE  \\
HD\,115363       & HIP 64896   &  305.88  &   -0.97  &   7.82  &	3282	    &	-55  & 290 & WE      & -35.3 &  0.0 & POP/VE   \\
ROA\,5701        &             &  309.24  &  +15.05  &  13.16  &  4800$^{7}$        &  1246  &  50 & --      & -46.5 &  0.0 & FER2/VE  \\
HD\,120908       &             &  312.25  &   +8.37  &   5.88  &   338              &    49  & 370 & --      &  -4.3 &  0.0 & S07/VE   \\    
HD\,480          &             &  319.45  &  -65.58  &   7.03  &   469              &   427  & 530 & --      &  -1.0 &  0.0 & S07/VE   \\
HD\,142919       &             &  328.43  &   -0.76  &   6.10  &        268         &    -4  & 500 & WE      & -3.2  &  0.0 & S07/VE   \\
HD\,186837       &             &  335.85  &  -30.57  &   6.20  &        329         &  -167  & 620 & WE      & -2.4  &  0.0 & S07/VE   \\  
IRAS\,17311      &             &  341.41  &   -9.04  &  11.4   &   1100$^{8}$       &  -174  &  55 & --      &  -9.5 &  0.0 & FER1/VE  \\
HD\,163758       &  SAO 209560 &  355.36  &   -6.10  &   7.32  & 4103	            &  -436  & 550 & --      & -16.1 &  0.0 & POP/VE   \\
HD\,163745       &             &  350.56  &   -8.79  &   6.50  & 2189               &   335  & 620 & --      & -11.9 &  0.0 & S07/VE   \\
BD+09\, 2860     &             &  353.04  &  +63.21  &  11.27  &  533$^{10}$        &   475  & 250 & --      &  -0.4 &  0.0 & S07/LD   \\
HD\,177566       &             &  355.55  &  -20.42  &  10.17  & 1100$^{9}$         &  -383  & 120 & --      &-190.2 &  0.0 & FER1/VE  \\  
CD-41\,13967     &             &  359.28  &  -33.50  &   9.5   & 3500$^{11}$        & -1900  &  80 & --      &  -1.2 &  0.0 & FER1/VE  \\
\hline
\end{tabular}
\begin{itemize}
\item[] Reference codes: 
(1)~Hoekzema, Lamers \& van Genderen (1993),
(1)~Smartt, Dufton \& Lennon (1997),
(2)~Gonz\'{a}lez et al. (2006)
(3)~Carney et al. (1994).
(4)~Hoekzema, Lamers \& van Genderen (1993),
(5)~Smoker et al. (2003),
(6)~Quin \& Lamers (1992)
(7)~Kinman et al. (2000),
(8)~Laird, Carney \& Latham (1988), 
(9)~Zsarg\'{o} et al. (2003),
(10)~Beers et al. (2000), 
(11)~McCarthy et al. (1991),
(12) From parallax.
(13) From RR-Lyrae calibration and magnitude. 
\end{itemize}
\normalsize
\end{center}
\end{table*}

\begin{figure*}
\setcounter{figure}{0}
\epsfig{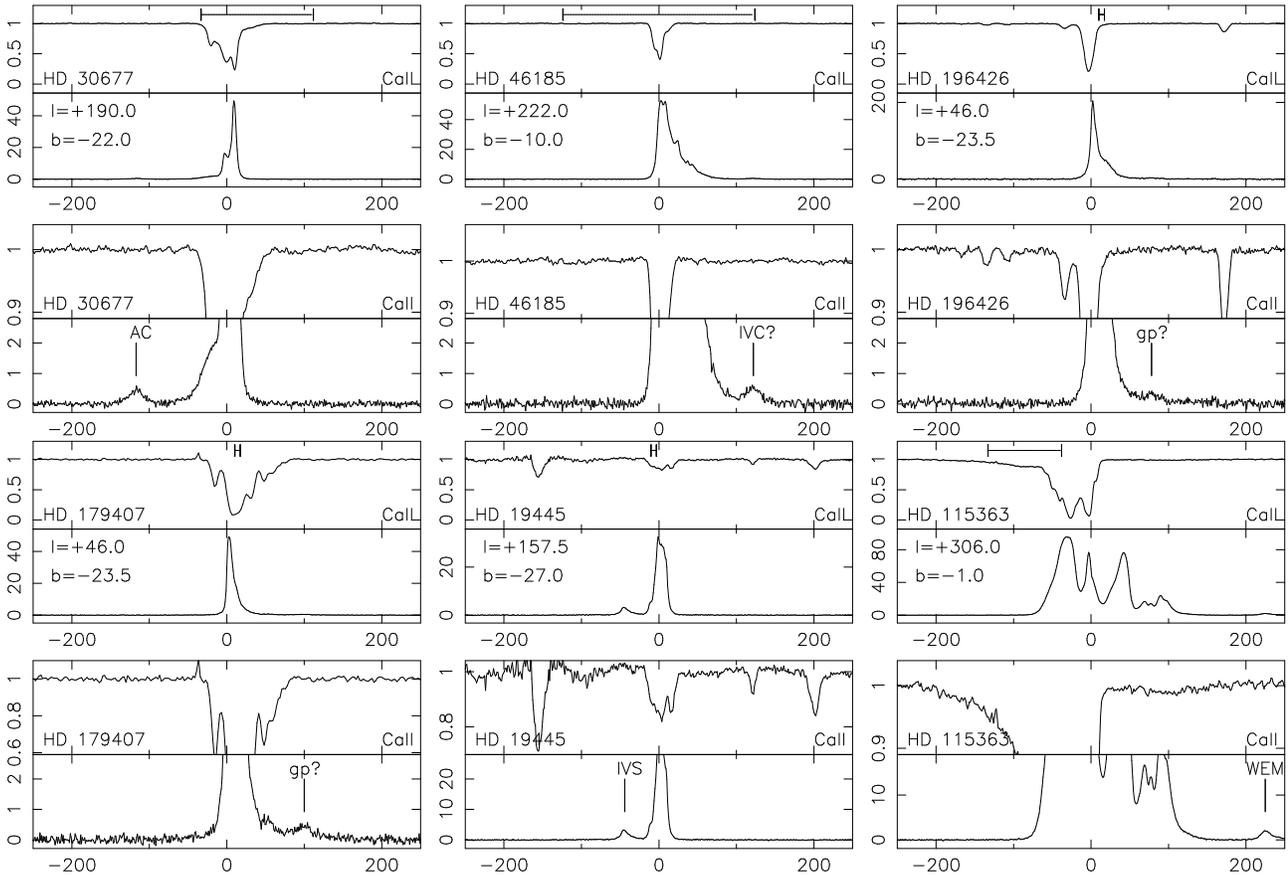}
\caption[]{Optical Ca\,{\sc ii} K and 21-cm H\,{\sc i} spectra towards early-type stars 
for which a lower distance limit towards an IHVC has been determined. Two plots are shown per sightline in 
order to emphasise weak features. Further details are given in the text.}
\label{SpectraIHVCs}
\end{figure*}

\section{Results}
\label{results}

In this section we discuss those cases where the stellar sightlines intersect 
with known IHVCs, and hence determine improved distance estimates towards a handful of objects. 

Fig. \ref{SpectraIHVCs} shows the optical and H\,{\sc i} spectra towards the sightlines where 
a distance limit has been determined towards an IHVC. Fig. \ref{SpectraAll} (available online) shows the remaining sightlines. 
Two plots are shown for each sightline in order to emphasise both weak and strong features. The majority of 
the optical spectra are in the Ca\,{\sc ii} K line; where this was not available the Ti\,{\sc ii} 
line is shown. The horizontal line at the top of the first of the optical plots shows the 
extent of the full width half maximum of the stellar line. In most cases, these lines are wide, 
hence there is no possibility that stellar lines could be misidentified as interstellar features, 
which tend to be much narrower. If the stellar lines have a FWHM exceeding $\sim$100 km\,s$^{-1}$ they 
were removed in the normalisation process to facilitate visualisation of the interstellar 
lines in all cases apart from HD numbers \,115363, 136239 and 142758 where too much 
overlap of stellar and interstellar components occur.

\subsection{Methodology of estimating distances to IHVCs}

The method of estimating distances to IHVCs is discussed fully 
in Schwarz, Wakker \& van Woerden (1995). For an upper distance limit, detection of optical absorption, 
in association with an H\,{\sc i} detection, is sufficient to provide 
an upper distance limit, being the distance of the stellar probe. 
Lower-distance limits are more problematic. A {\em firm} lower distance limit 
can only be set if no optical absorption is seen at a sufficient S/N ratio, 
the abundance of the optical element is known (generally from observations of the 
same part of the complex towards QSOs), and the H\,{\sc i} column density is accurately 
defined using a pencil beam. For the current sample, the chemical abundance of the IHVC 
is often not known, and the observations in H\,{\sc i} only have a spatial resolution 
of 0.5$^{\circ}$, which means that care must be taken in ascribing a lack of 
optical absorption as due to the stellar probe being closer than the IHVC. 
However, these factors are somewhat ameliorated by the fact that the optical 
spectra have high S/N, frequently being $>$500 per resolution element and with a median 
of 410 in the sightlines with a detected IHVC.

\subsection{Distance limits towards individual complexes}
\label{distlim}

A number of the current sightlines either intersect with known IHVC complexes, 
or have gas present at IHVC velocities in the Villa-Elisa or Leiden-Dwingeloo H\,{\sc i} spectra. These 
cases are discussed below, and lower distance estimates towards five IHVCs are determined. 
Table \ref{t_distIHVCs} summarizes these cases. Columns 1--6 gives the star name, stellar distance, 
previous IHVC distance limit, IHVC complex, observed H\,{\sc i} velocity and corresponding log of the H\,{\sc i} column 
density. Columns 7--8  give the previously-known abundance in Ca\,{\sc ii} taken from Wakker (2001) and limiting 
5$\sigma$ Ca\,{\sc ii} column density estimated from the current spectra. This was 
derived using the observed S/N ratio and instrumental resolution, 
assuming the the optically thin approximation. Finally, column 9 gives the {\em predicted} 
Ca\,{\sc ii} column density derived by subtracting the previously-known Ca\,{\sc ii} abundance from the log 
of the observed H\,{\sc i} column density. Where this predicted value is much higher than the limiting 5$\sigma$ Ca\,{\sc ii} 
column density a non-detection is interpreted as the cloud lying further away than the stellar probe. Individual 
complexes are discussed below.  

%
%
%
%

\begin{table*}
\begin{center}
\small
\caption[]{IHVC sightlines where H\,{\sc i} is detected at intermediate or high velocities. Complex WEM is in the same part of 
the sky as complex WE of Wakker (2001), but at higher velocities. See Sect \ref{distlim} for details.}
\label{t_distIHVCs}
\begin{tabular}{lrrrrrrrrr}
\hline  
    Star        &    $d$   & $d_{\rm IHVC}^{\rm prev}$ &   IHVC    &    $v_{\rm IHVC}$(H\,{\sc i})   &  log($N_{\rm IHVC}$(H\,{\sc i}))   &  $A_{\rm IHVC}^{\rm prev}$(Ca\,{\sc ii})  &  log($N_{\rm lim}$(Ca\,{\sc ii})) &  log($N_{\rm pred}$(Ca\,{\sc ii}))       \\
                &   (pc)   &     (pc)                  &  complex  &    km\,s$^{-1}$      &  (log(cm$^{-2}$))       &  (log(cm$^{-2}$))              &  (log(cm$^{-2}$))                 &  (log(cm$^{-2}$))                        \\
\hline
 HD\,196426     &   700    &       800-4300            &     gp    &      +78             &	     19.06           &       -7.42                   &           10.20                     &          11.65                        \\
 HD\,179407     &  7600    &          "                &     gp    &      +50             &	     18.49           &       -7.42                   &           10.60                     &          11.07                        \\
      "         &    "     &          "                &     gp    &      +97             &          18.49           &       -7.42                   &           10.60                     &          11.07                        \\
 HD\,19445      &    39    &          --               &    IVS    &      -45             &          19.49           &       -7.88                   &           10.46                     &          11.61                        \\
      "         &    "     &          --               &    IVS    &      -40             &          19.71           &       -7.88                   &           10.46                     &          11.82                        \\
 HD\,30677      &  2700    &       $>$400              &   ACII    &     -117             &          19.24           &      $<$-8.39                 &            9.82                     &             --                        \\
 HD\,115363     &  3200    &          --               &    WEM    &     +224             &          19.71           &         --                    &            9.99                     &             --                        \\
      "         &    "     &          --               &    WEM    &     +240             &          19.30           &         --                    &            9.99                     &             --                        \\
 HD\,46185      &  2900    &          --               &  Other    &     +122             &          19.09           &         --                    &            9.71                     &             --                        \\             
\hline
\end{tabular}
\end{center}
\end{table*}

\subsubsection{Complex gp IVC}

Complex gp is a positive-velocity IVC lying in the direction of the globular 
cluster M\,15, which has previously been studied in infrared, optical, H$\alpha$ and 
H\,{\sc i} by Smoker et al. (2002). The previously-existing distance limit was
0.8--4.3 kpc (Wakker 2001 and references therein) with an uncertain lower distance limit 
of 2.0-kpc (Smoker et al. 2006). The Complex has LSR velocities 
of $\sim$+60 to +90 km\,s$^{-1}$. In our current sample, the
star HD\,188294 lies towards this Complex, but only has a distance 
of 212-pc and no H\,{\sc i} is detected for this sightline due to it being in a ``hole" in 
the Complex. Additionally, HD\,196426 ($l,b$=45.81$^{\circ}$,--23.32$^{\circ}$) lies 
towards Complex gp, and weak H\,{\sc i} is observed in emission in the Leiden-Dwingeloo spectrum, 
with a LSR velocity +78$\pm$1 km\,s$^{-1}$, a FWHM of 24$\pm$2 km\,s$^{-1}$, peak brightness temperature 
$T_{\rm B}$=0.25$\pm$0.05 K and brightness temperature integral of 6.5$\pm$1.0 K km\,s$^{-1}$, 
corresponding to an H\,{\sc i} column density of 1.1$\pm$0.2$\times$10$^{19}$ cm$^{-2}$. Although 
weak, this should have been detected in our UVES spectrum which has a S/N = 500 per 
resolution element. The star has a distance of 700-pc, which is similar to the distances for 
previous objects towards which there were non-detections. In Complex gp we also observed 
HD\,179407 ($l,b$=24.02$^{\circ}$,--10.4$^{\circ}$, distance=7600-pc) at a S/N pixel$^{-1}$  
of 120 in Ca\,{\sc ii} K. At the current position, there are two weak H\,{\sc i} velocity features, 
at $v$=+50$\pm$1 and $v$=+97$\pm$1 km\,s$^{-1}$ with FWHM values of 26$\pm$2 and 42$\pm$4 km\,s$^{-1}$
and brightness temperature integrals of 1.7$\pm$0.2 and 1.7$\pm$0.2 K km\,s$^{-1}$ respectively, corresponding 
to column densities of $\sim$3$\times$10$^{18}$cm$^{-2}$. There is obvious detection of Ca\,{\sc ii} in the 
+50 km\,s$^{-1}$ feature (as in the Ca\,{\sc ii} spectrum of Sembach \& Danks 1994), 
but no detection of the v$=+97$ km\,s$^{-1}$ feature, perhaps due to clumpiness in the H\,{\sc i} or ionisation issues; 
a higher S/N Ca\,{\sc ii} spectrum would be useful.  
Given the weak nature of both H\,{\sc i} features 
a higher spatial-resolution and sensitivity H\,{\sc i} spectrum would be useful at this position although in 
any case the star lies at a distance exceeding the current upper limit of the cloud. Although HD\, 179407 was 
also observed in the  
FUSE spectrum by Zsargo et al. (2003), the presence of a complex stellar continuum meant that 
no interstellar O\,{\sc vi} was observed. Finally, although EC 20485-2420 lies in the general direction of this complex,
no H\,{\sc i} is obvious in the Villa-Elisa spectrum.

\subsubsection{IV South}

IV South is a group of IVCs that extend over much of the southern sky, with velocities of 
$\sim$--85 to --45 km\,s$^{-1}$. Towards HD\,19445 ($l,b$=157.48$^{\circ}$,--27.20$^{\circ}$), 
no IV absorption is seen in the Ca\,{\sc ii} spectrum at a S/N of 280 per  resolution element, thus placing 
a rather uninteresting firm lower distance limit of 39-pc to this part of the IVC that has two components 
with $v$=--45$\pm$0.5 km\,s$^{-1}$, --40.2$\pm$0.5 km\,s$^{-1}$, FWHM values of 8$\pm$1 km\,s$^{-1}$ 
and 22$\pm$2 km\,s$^{-1}$, peak T$_{B}$ values of of 2.1$\pm$0.2 K and  1.2$\pm$0.2 K and 
brightness temperature integral of 17$\pm$2 K km\,s$^{-1}$ and 28$\pm$3 K km\,s$^{-1}$. The combined  
H\,{\sc i} column density in these two features is $\sim$8$\times$10$^{19}$ cm$^{-2}$ which should 
have been easily detected in the current optical spectrum if the cloud were closer than the star. 

\subsubsection{Complex K}

Complex K is a Northern-Hemisphere cloud with LSR velocities ranging from --65 to --95 km\,s$^{-1}$. 
Its previous distance bracket was $\sim$700--6800-pc (Smoker et al. 2006 and refs. therein). One of 
our sightlines towards G169-28 ($l,b$=41.83$^{\circ}$,+36.06$^{\circ}$) lies in the general direction 
of Complex K, but no H\,{\sc i} emission is visible in the Leiden-Dwingeloo spectrum and there are many 
stellar lines. Thus the current observations do not add anything to our knowledge of this IVC. 

\subsubsection{Anti-centre HVCs}

Seven of our sightlines lie in the region of the Anti-Centre HVC (Fig. 
9 of Wakker 2001). No upper distance limit is available for this HVC and 
the previous lower-distance limit towards Cloud ACI is only 0.4-kpc (Tamanaha 1996). 
We only detect H\,{\sc i} at high velocity towards one of the current 
sightlines which lies towards ACII, namely HD\,30677 at a velocity 
of --117$\pm$1 km\,s$^{-1}$, peak brightness 
temperature of 0.40$\pm$0.04 K, FWHM of 23$\pm$2 km\,s$^{-1}$ 
and integrated brightness temperature of 9.5$\pm$1.0 K km\,s$^{-1}$, 
corresponding to an HVC column density of 1.7$\pm$0.2$\times$10$^{19}$ 
cm$^{-2}$. Assuming that the HVC has a similar abundance to the 
relation from Wakker \& Mathis (2000), we would expect a corresponding
column density log(Ca\,{\sc ii} cm$^{-2}$)=11.64. However, 
no corresponding optical absorption is detected in our 
Ca\,{\sc ii} K spectrum, which has a S/N = 430 pixel$^{-1}$ or 
610 per resolution element. Assuming that the cloud is optically
thin in Ca, a 5$\sigma$ detection, f = 0.634 for the Ca\,{\sc ii} K transition
and instrumental resolution 
of 0.05\AA, the limiting column density observable with the current spectrum is 
log(Ca\,{\sc ii} cm$^{-2}$)=9.82, more than a factor 60 
lower than predicted from 
the H\,{\sc i} profile. Hence the current observations put a firm lower distance 
limit of 2.7-kpc towards complex ACII, assuming that the H\,{\sc i} observed 
in the Villa-Elisa survey reflects that in the pencil beam towards HD\,30677.

\subsubsection{Complex WE/WEM HVC}

Complex WE is a group of small HVCs centred on ($l,b$)$\sim$(320$^{\circ}$,0$^{\circ}$),  
first detected by Mathewson, Cleary \& Murray (1974) and  mapped in H\,{\sc i} by Morras 
(1982). Parts of it lie in the same region of the sky as two large low-velocity H\,{\sc i} shells 
in the direction of the Coalsack nebula described by McLure-Griffiths et al. (2001). 
At $b\sim$0$^{\circ}$ latitude the predicted values of Galactic rotation at 
$l\sim$320$^{\circ}$ are from $\sim$--120 to +70 km\,s$^{-1}$, falling to $\sim$--100 to 0 km\,s$^{-1}$ 
at $b\sim$--15$^{\circ}$. Towards HD\,156359 ($l,b$=328.68$^{\circ}$, --14.52$^{\circ}$), 
Sembach et al. (1991) found optical absorption at $\sim$+110 km\,s$^{-1}$, putting an upper 
distance limit of 12.8 kpc. Eighteen of our sightlines lie 
within the general area of WE as defined in Fig. 11 of Wakker (2001). One 
of the sample stars HD\, 115363 ($l,b$=306.0$^{\circ}$,--1.0$^{\circ}$ with spectroscopic distance=3.2-kpc) 
has HVC gas detected with two components at +224.5$\pm$3.0 km\,s$^{-1}$, 
+240.0$\pm$5.0 km\,s$^{-1}$, velocity widths 14.4$\pm$0.8 km\,s$^{-1}$ and 
19.2$\pm$2.4 km\,s$^{-1}$, peak brightness temperatures of 1.8$\pm$0.06 K and 
1.1$\pm$0.1 km\,s$^{-1}$ and brightness temperature integrals of 
28.3$\pm$1.0 K km\,s$^{-1}$ and 11.1$\pm$0.8 K km\,s$^{-1}$ which correspond 
to H\,{\sc i} column densities of 5.2$\pm$ 0.2$\times$10$^{19}$ cm$^{-2}$ and
2.0$\pm$0.1$\times$10$^{19}$ cm$^{-2}$. There is no Ca\,{\sc ii} K absorption present 
in the spectrum, which has a S/N = 410 per resolution element. This HVC is 
probably associated with the clouds defined by Putman (2000) as the Leading Arm: the 
counterpart of the Magellanic Stream, as projected on the sky, between the Magellanic Clouds and the 
Galactic Plane. These data hence set an unsurprising lower limit of 3.2-kpc towards this HVC that is probably 
related to the Magellanic System, using our distance estimated spectroscopically . If we assume that HD\,115363 is a 
part of the Centaurus OB1 association, its distance is slightly closer at 2.5-kpc (McClure-Griffiths et al. 2001 and refs. therein).
Finally we note that this HVC appears to be a different set of clouds to the lower-velocity and more negative 
galactic-latitude clouds described in Wakker (2001) and observed by Sembach et al. (1991), hence in the 
current paper it is named WEM due to its possible association with the Magellanic system.

\subsubsection{Other IVCs}

In the line-of-sight towards HD\,46185 ($l,b$=222.0$^{\circ}$,--10.1$^{\circ}$), H\,{\sc i}
emission is detected at +122$\pm$2 km\,s$^{-1}$, with a peak brightness temperature of 0.35 K, FWHM of 
17$\pm$3 km\,s$^{-1}$ and brightness temperature integral of 6.7$\pm$0.7 K km\,s$^{-1}$,
corresponding to an H\,{\sc i} column density of 1.2$\pm$0.1$\times$10$^{19}$ cm$^{-2}$. 
Normal Galactic rotation predicts velocities of upto $\sim$+97 km\,s$^{-1}$ in this part of the sky, 
so the deviation velocity is only $\sim$ 25 km\,s$^{-1}$ and the cloud many not be an IVC. 
Assuming that the cloud has a similar abundance 
to the relation from Wakker \& Mathis (2000), we would expect a column density
log(Ca\,{\sc ii} K cm$^{-2}$)=11.59. However, no corresponding optical absorption
is detected in our Ca\,{\sc ii} K spectrum, which has a S/N = 550 
pixel$^{-1}$ or 780 per resolution element. The 5$\sigma$ limiting column density observable 
with the current spectrum is 
log(Ca\,{\sc ii} cm$^{-2}$)=9.71, a factor 75 lower than predicted from
the H\,{\sc i} profile. Hence the current observations put a firm lower 
distance limit of 2.9-kpc towards this parcel of gas that lies within $\sim$20$^{\circ}$ 
of the Anti-Centre Shell (Fig. 8 of Wakker 2001) but is at different velocities 
and probably unrelated.

\subsection{IHVCs detected in Ca\,{\sc ii} absorption}

A number of sightlines were already flagged in Paper {\sc i} as having IHVC components detected in 
the optical spectra. These include the Wolf Rayet stars HD\,94910 and HD\,163758 and the sightline 
HD\,72067 which lies towards the Vela Supernova remnant. No H\,{\sc i} is detected towards any of 
these sightlines. In the first two cases this implies the presence of circumstellar lines and 
in the latter case lines within the SN remnant.  Similarly, towards HD\,171432 many IVCs are detected 
in the optical. This sightline lies towards the Scutum Supershell mapped 
in H\,{\sc i} by Callaway et al. (2000) and with a distance of $\sim$3000-pc. Although towards 
HD\,171432 there is a dearth of H\,{\sc i} detected in the Callaway maps, there {\em is} H\,{\sc i} 
in our H\,{\sc i} spectrum up to a velocity of $\sim$+90 km\,s$^{-1}$, coincident with our  
detections of Ca\,{\sc ii}. No H\,{\sc i} is seen in our highest-velocity Ca\,{\sc ii} component 
of $\sim$+120 km\,s$^{-1}$, perhaps due to S/N limitations. The detections in Ca\,{\sc ii} and 
H\,{\sc i} are consistent with the supershell being closer than our stellar distance of 
$\sim$4000 pc and with the previous observations, but add nothing to the distance bracket.

\section{Discussion}
\label{discussion}

Table \ref{t_MyDist} gives a summary of the distance limits to IHVCs set by the current observations,
plus existing limits to the clouds where available. Particularly interesting is the improved lower limit towards part of 
the Anti-Centre complex ACII which has firm lower-distance limit of $>$2.7-kpc. This compares with the indirect 
distance estimate of a part of the complex at $l\sim$60$^{\circ}$,$b\sim$--45$^{\circ}$ derived from morphological 
and kinematical arguments of $\sim$ 4-kpc (Peek et al. 2007), and an H$\alpha$ estimated distance of between 8 and 20-kpc 
(Weiner et al. 2001) which is based upon the observed ionisation being caused by the Galactic 
radiation field. Although a big improvement on the previous lower-distance limit of $\sim$0.4-kpc (Tamanaha 1996), 
the current observations cannot discriminate between the indirectly-estimated distances and clearly 
searches for more distant probe stars in this part of the sky would be useful. Other less interesting results 
are the consolidation of the lower-distance limit towards complex gp and the first lower distance limit towards the 
WEM complex. The $z$-distance of the former IVC is now 
constrained to $\sim$300-1700-pc which compares to the H\,{\sc i} scaleheight of $<$200-pc at Galactocentric 
radii of $<$10-kpc (Narayan, Saha \& Jog 2005). Further progress on this sightline should involve performing obtaining 
a high-resolution spectrum of the star HD\,357657 and associated model atmosphere calculation and abundance analysis.   
Although Smoker et al. (2006) estimated a distance of $\sim$2.0-kpc for this object on the line of sight to Complex gp 
and found no associated Ca\,{\sc ii} absorption, the distance of the star remains uncertain. {\em If} a firm lower distance 
limit of 2-kpc were confirmed, cloud parameters such as the cloudlet sizes, cloud electron density, fractional H\,{\sc i} 
to H\,{\sc ii} ratios and ionizing radiation field could be better constrained (c.f. Smoker et al. 2002), and the 
position of the cloud relative to the H\,{\sc i} disc of the Galaxy confirmed.  

Finally, the lower distance limit of 3.2-kpc towards HVC WEM is consistent with both a Magellanic origin as proposed for example 
by Putman (2000), or a 'classical' high velocity cloud. H\,{\sc i} synthesis mapping towards other HVCs in this part of the sky 
(e.g. Bekhti et al. 2006) have provided evidence from cloud structure and linewidths of distances of $\sim$10--60-kpc, 
consistent with a Magellanic origin, and the same observations could be performed for the present sightline in order to 
obtain an indirect distance estimate, perhaps in conjunction with H$\alpha$ mapping. However, in the absence of early-type 
stars present in the leading arm as present in the Magellanic Bridge (Rolleston et al. 1999), obtaining a firm upper 
distance limit will be difficult although perhaps possible due to the offset in velocity from the stellar 
and interstellar Ca\,{\sc ii} K lines (c.f. Smoker et al. 2002).
 
\begin{table}
\begin{center}
\caption[]{Distance limits and probe stars towards the IHVCs studied in this paper.}
\label{t_MyDist}
\begin{tabular}{lrrrrrr} 
\hline
IHVC       & ($l,b$)    & $v_{\rm IHVC}$(H\,{\sc i}) &  Probes     & $D_{\rm IHVC}$     \\
           &  (deg.)    &  km\,s$^{-1}$              &             &  (pc)              \\	   
\hline
gp         &  46,--23   &         +78     	     & HD\,196426  & 800-4300$^{1,2}$   \\
IVS        & 157,--27   &     --45, --40             & HD\,19445   &  $>$ 39$^{1}$      \\
ACII       & 190,--22   &    --117                   & HD\,30677   &  $>$ 2700$^{1}$    \\
WEM        & 306,--1    &  +224, +240                & HD\,115363  &  $>$ 3200$^{1}$    \\
Other      & 222,--10   &    +122                    & HD\,46185   &  $>$ 2900$^{1}$    \\
\hline
\end{tabular}
\item[] Reference codes: 
(1) This paper,
(2) Little et al. (1994).
\end{center}
\end{table}

\section{Summary}
\label{summary}

We have correlated optical spectra in the Ca\,{\sc ii} K and Ti\,{\sc ii} lines
observed towards early-type stars in the POP Survey, plus other optical data, with 
21-cm H\,{\sc i} spectra taken from the Villa-Elisa and Leiden-Dwingeloo Surveys, in 
order to determine the distances to Intermediate and High Velocity Clouds. The
lack of Ca\,{\sc ii} K absorption at --117 km\,s$^{-1}$ towards HD\,30677 at a S/N 
ratio of $\sim$610 has set a firm lower distance limit towards Anti-Centre cloud ACII which 
previously had a lower distance limit of 0.4-kpc. Likewise, towards HD\,46185 no Ca\,{\sc ii} K absorption 
at +122 km\,s$^{-1}$ is seen at a S/N ratio of $\sim$780, hence placing a lower distance limit 
of 2.9-kpc towards this gas that is perhaps an IVC. Towards Complex gp no Ca\,{\sc ii} K absorption is 
seen in the spectrum of HD\,196426 at a S/N of $\sim$500, reinforcing the assertion that this IVC lies 
at a distance exceeding 0.7-kpc. Likewise, towards the nearby star HD\,19445 at 39-pc in the line of 
sight to IV South no Ca\,{\sc ii} K absorption is seen setting a a firm but uninteresting distance limit towards
this part of the complex. Finally, no HV Ca\,{\sc ii} K absorption is seen in the stellar spectrum of HD\,115363 
at a S/N = 410, placing a lower distance of $\sim$3.2-kpc towards the HVC gas at velocity of $\sim$+224 km\,s$^{-1}$. 
This gas is in the same region of sky as the WE complex of Wakker (2001), but at higher velocities. If related to the 
Magellanic system (Putman 2000) then a distance limit of 3.2-kpc is not unexpected. 

A future paper will describe the use of new FEROS observations combined with UVES 
archive data to provide improved distance limits to complex EP, the Cohen Stream, IV South and the Anti-Centre shell. 
Concerning the POP data, future papers will investigate the neutral species of Ca\,{\sc i}, Fe\,{\sc i}, 
Na\,{\sc i} D and K\,{\sc i} as well as molecular line species CH, CH$^{+}$ and CN in order to 
better understand the local interstellar medium.

\section*{acknowledgements}                            \label{s_acknowledgments}

We would like to thank the staff of the Very Large Telescope, Paranal for 
the large amount work involved in producing the POP Survey 
(ESO DDT programme ID 266.D-5655(A), http://www.eso.org/uvespop). Especially 
due thanks are S. Bagnulo, R. Cabanac, E. Jehin, C. Ledoux and C. Melo. 
In addition, we are grateful to the staffs of Dwingeloo/Leiden and the Villa-Elisa Telescope 
for producing the H\,{\sc i} all sky surveys. JVS and HMAT thank the support staff at La Silla for their help 
with the FEROS observations. 
FPK is grateful to AWE Aldermaston for the award of a William Penney Fellowship. This research has made use 
of the {\sc simbad} Database, operated at CDS, Strasbourg, France. JVS acknowledges financial 
support from the Particle Physics and Astronomy Research Council with HMAT and IH thanking 
the Department of Education and Learning for Northern Ireland. JVS thanks M. Garc\'{i}a Mu\~niz, 
L. Salinas and I. Dino for discussions and to an anonymous referee for comments.


{}

\begin{figure*}
\epsfig{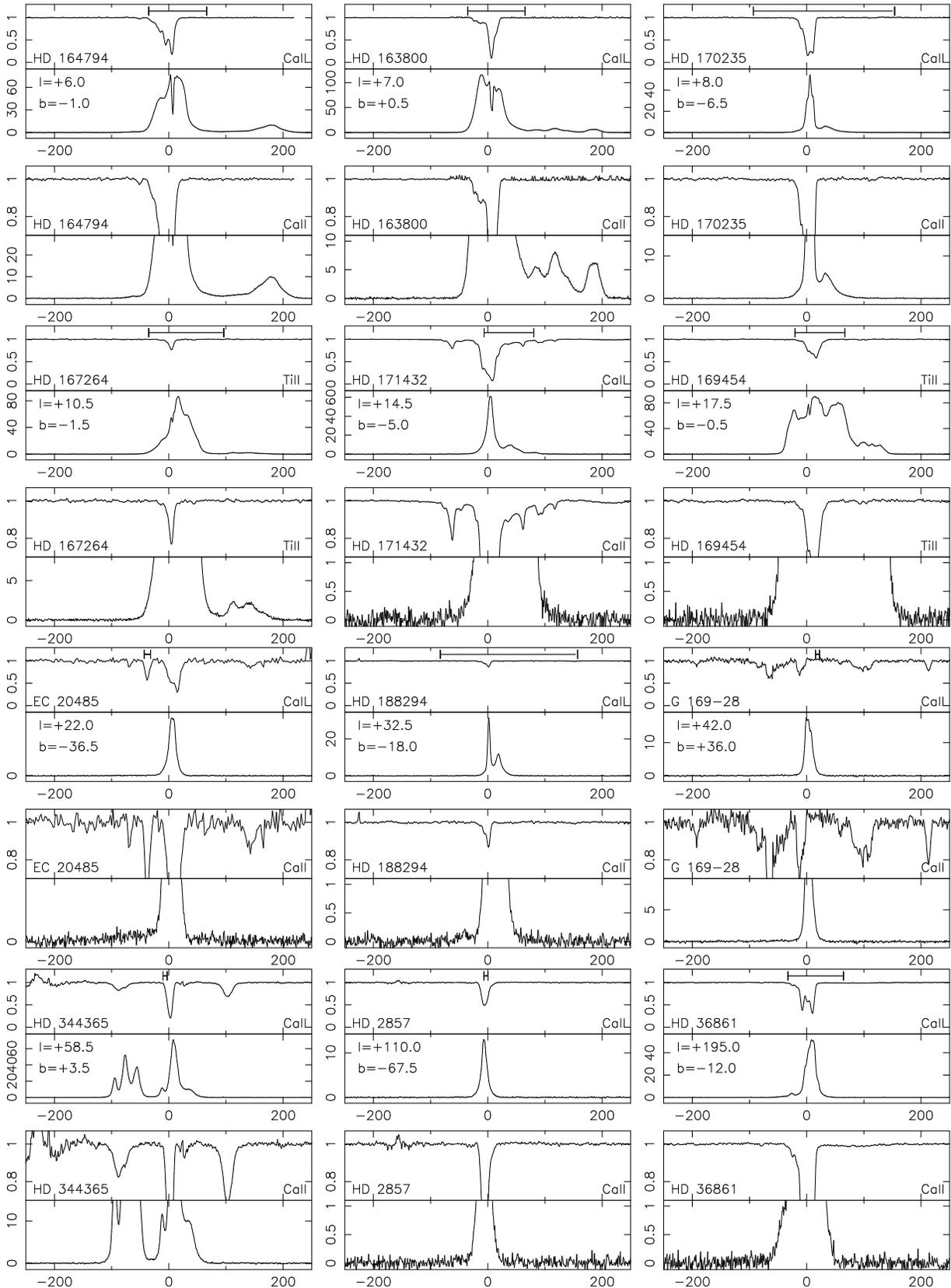}
\caption[]{Optical (Ca\,{\sc ii} K or Ti\,{\sc ii}) and 21-cm H\,{\sc i} spectra towards early-type stars. 
Two plots are shown per sightline in order to emphasise weak features. In the cases where the FWHM of the 
stellar profile exceeded $\sim$100 km\,s$^{-1}$ it has been removed in the normalisation process to emphasise the
interstellar line features. This affects the stars with HD numbers 480, 2857, 2913, 49131, 61429, 74966, 76131, 
90882, 100841, 115842, 122272. 145482, 156575, 163745, 186837, 188294, 344365, plus ROA 5701. For HD\,2857 some residuals are 
left by this process at $\sim$ --160 km\,s$^{-1}$.
}
\label{SpectraAll}
\end{figure*}

\begin{figure*}
\setcounter{figure}{1}
\epsfig{file=./Fig2_p2.eps,angle=0}
\caption[]{$ctd.$}
\end{figure*}

\begin{figure*}
\setcounter{figure}{1}
\epsfig{file=./Fig2_p3.eps,angle=0}
\caption[]{$ctd.$}
\end{figure*}

\begin{figure*}
\setcounter{figure}{1}
\epsfig{file=./Fig2_p4.eps,angle=0}
\caption[]{$ctd.$}
\end{figure*}

\begin{figure*}
\setcounter{figure}{1}
\epsfig{file=./Fig2_p5.eps,angle=0}
\caption[]{$ctd.$}
\end{figure*}

\begin{figure*}
\setcounter{figure}{1}
\epsfig{file=./Fig2_p6.eps,angle=0}
\caption[]{$ctd.$}
\end{figure*}

\begin{figure*}
\setcounter{figure}{1}
\epsfig{file=./Fig2_p7.eps,angle=0}
\caption[]{$ctd.$}
\end{figure*}

\begin{figure*}
\setcounter{figure}{1}
\epsfig{file=./Fig2_p8.eps,angle=0}
\caption[]{$ctd.$}
\end{figure*}

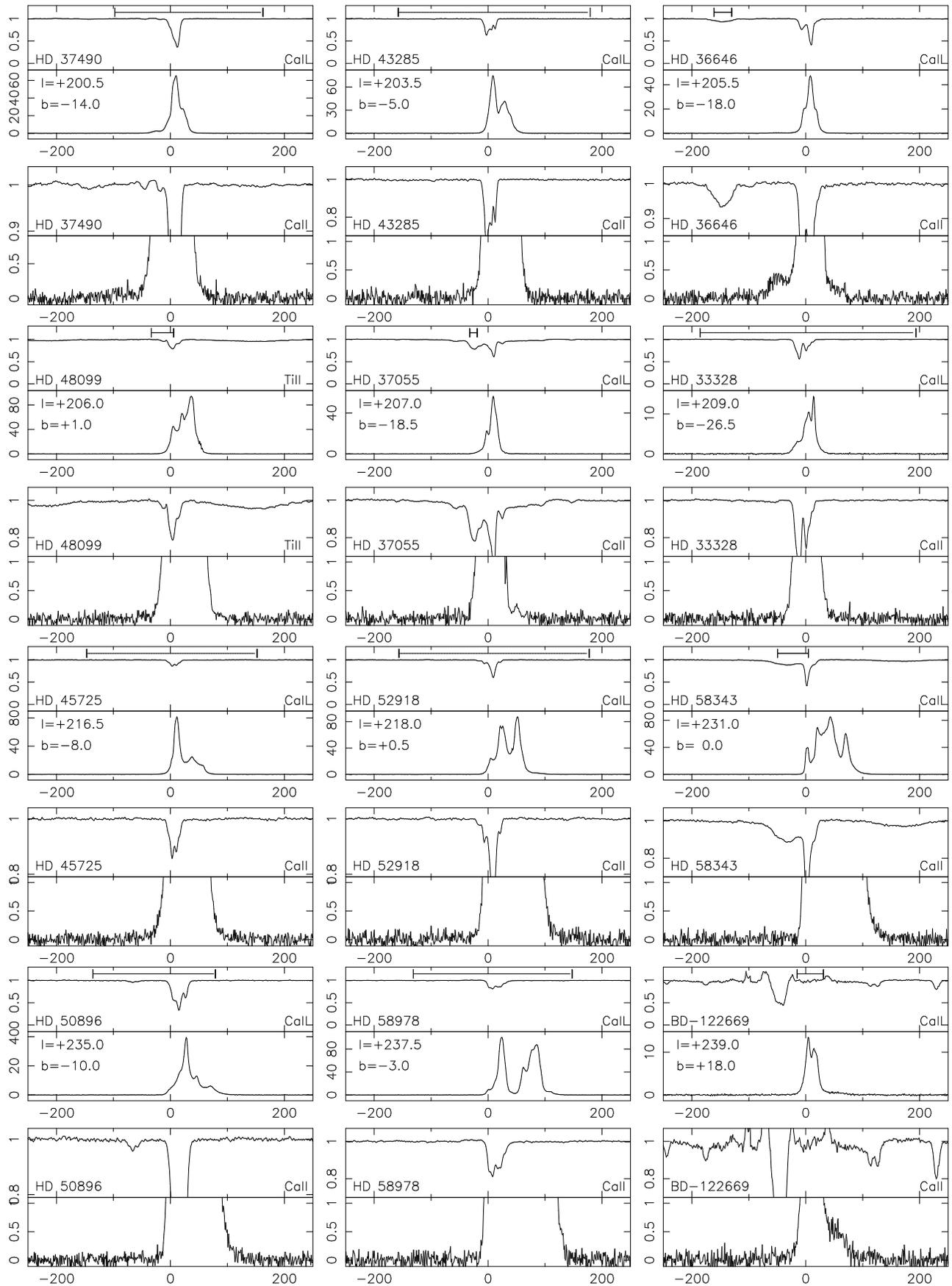
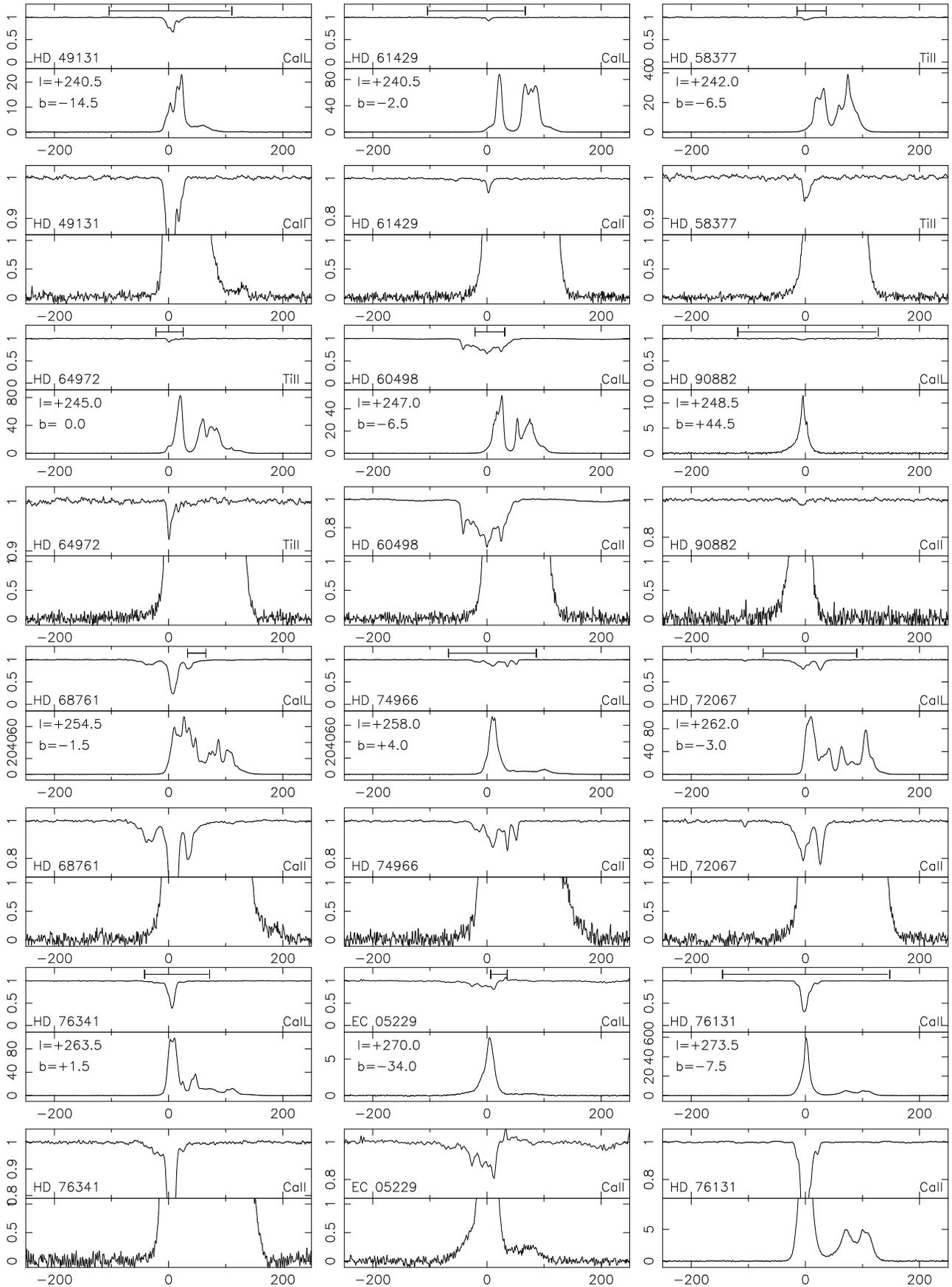
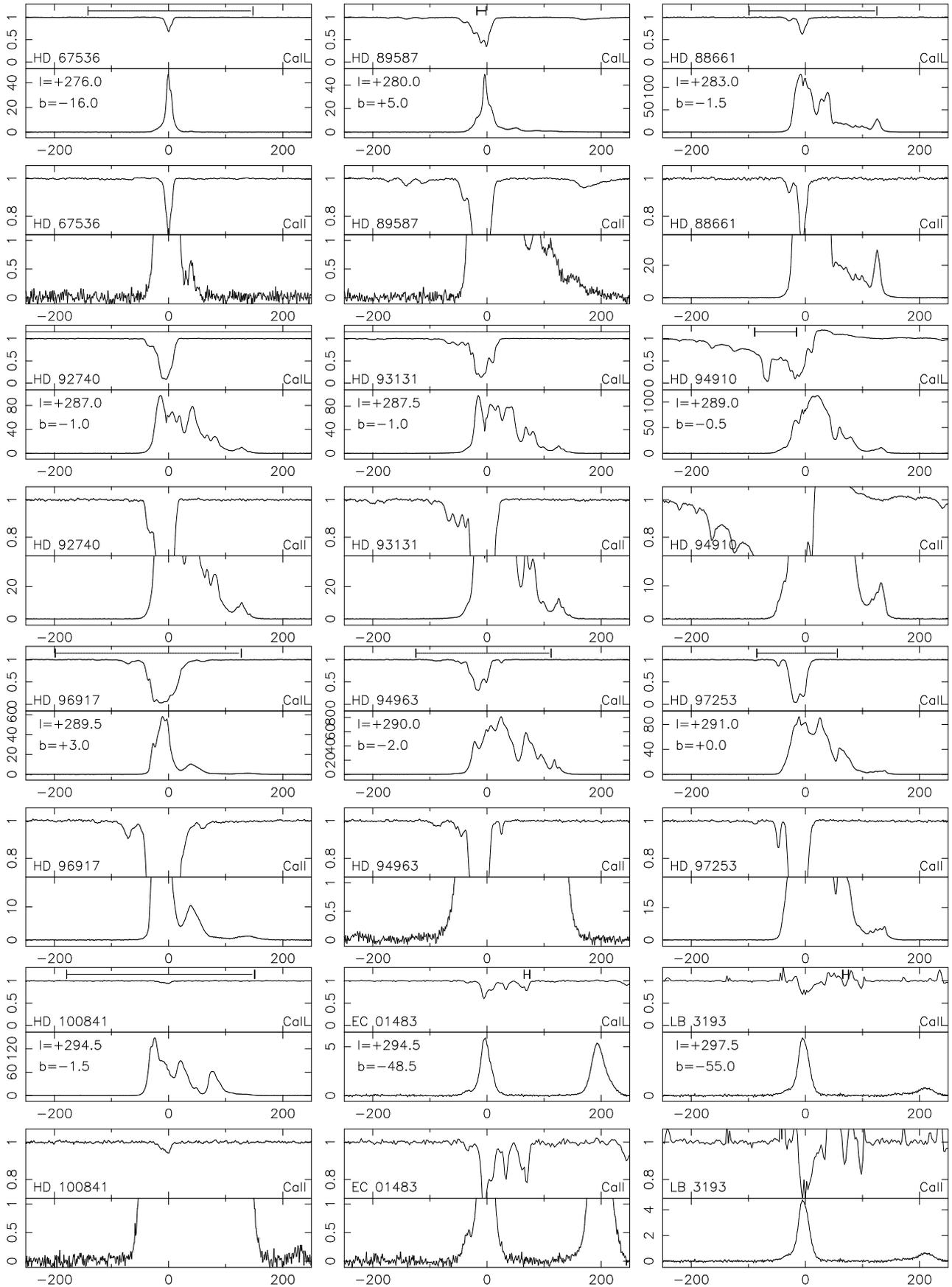
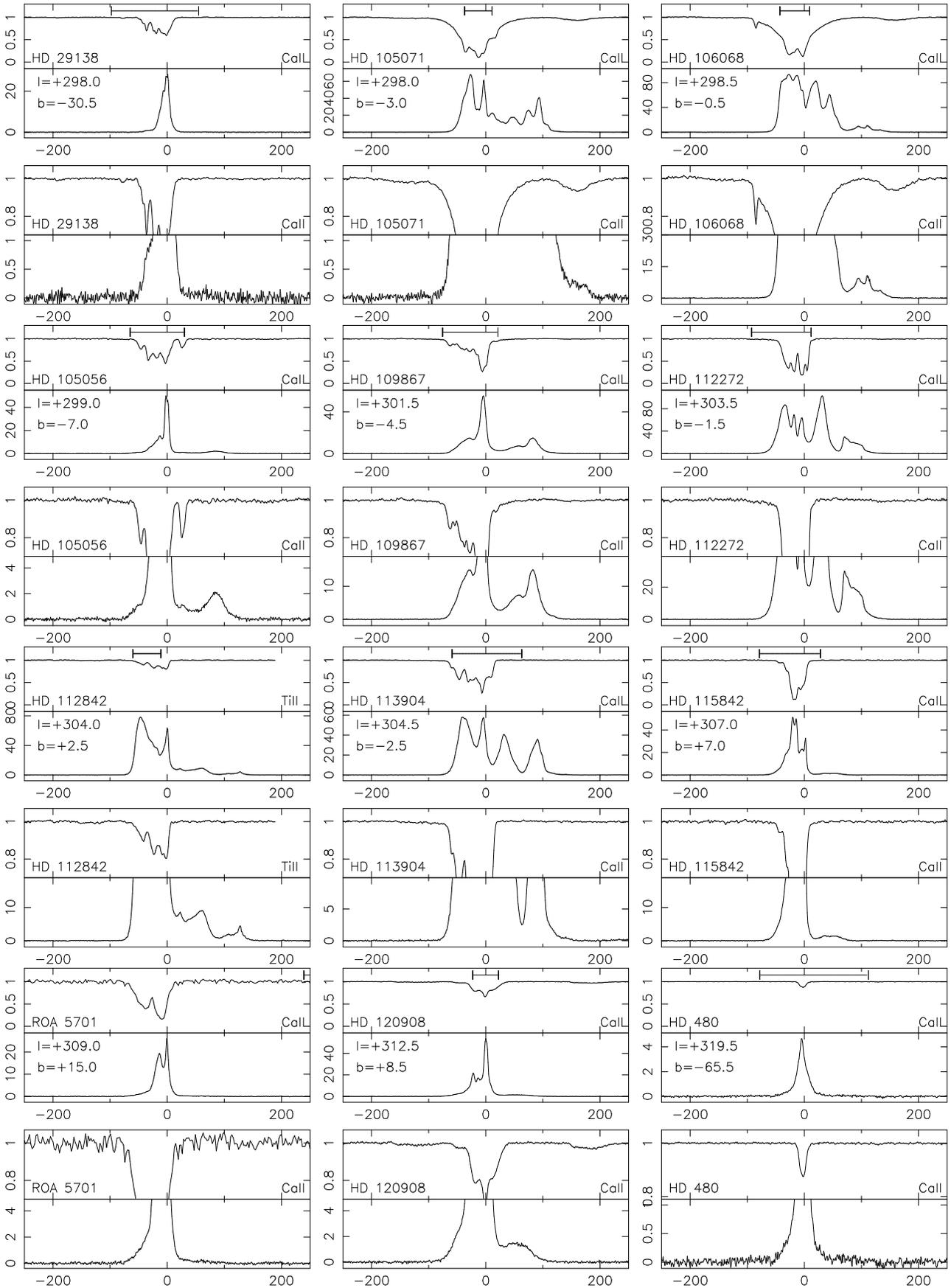
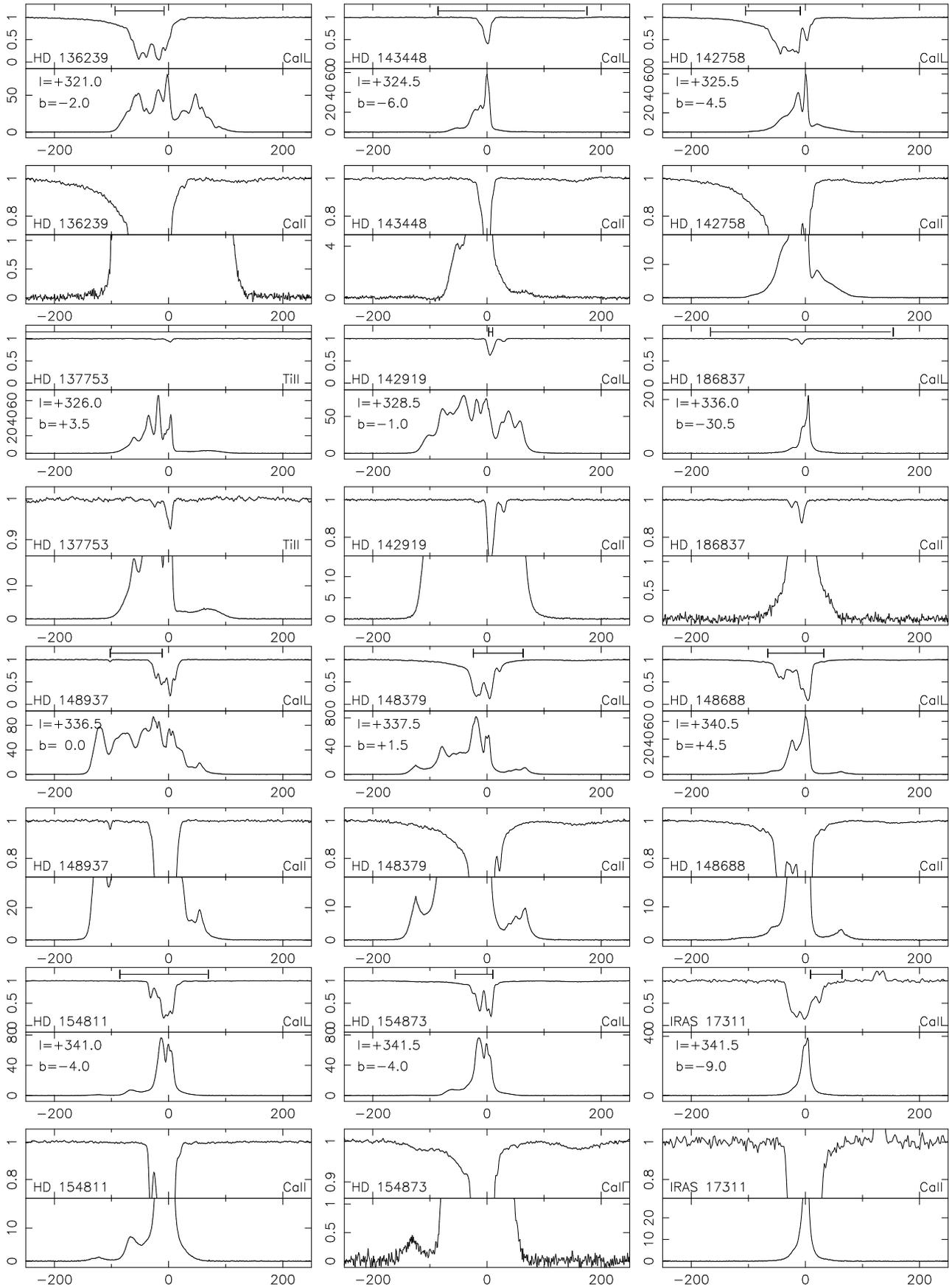
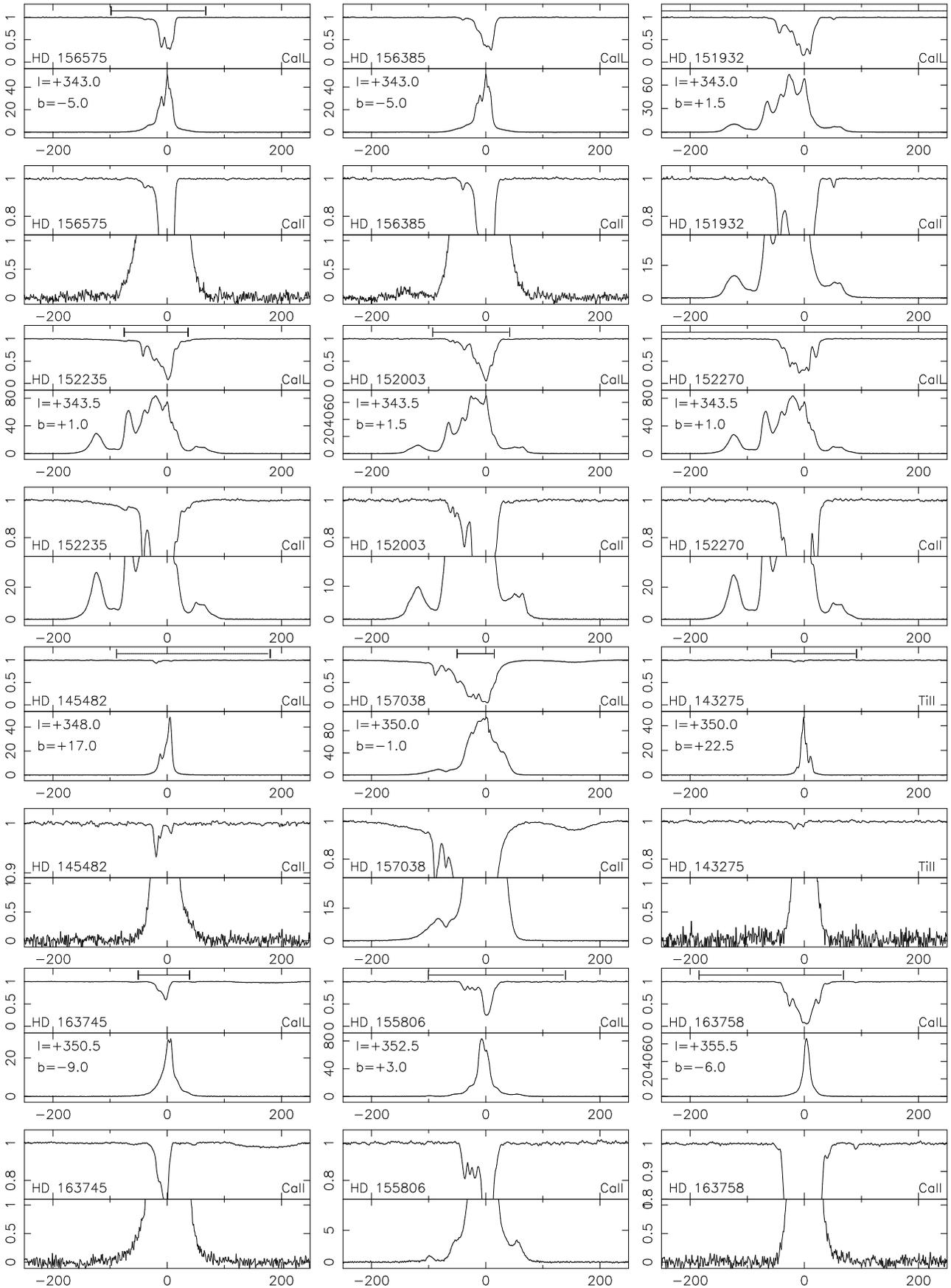
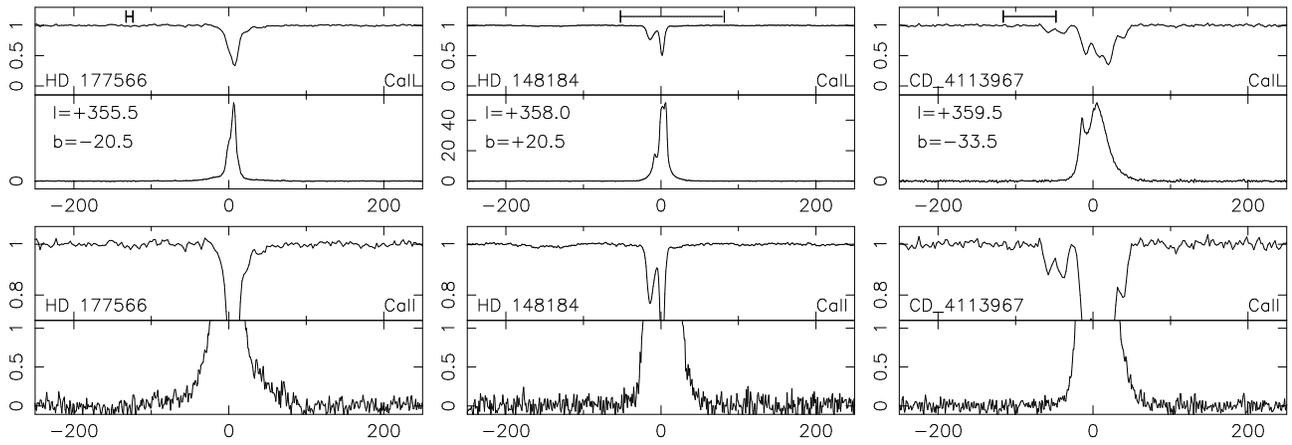
\end{document}